\documentclass[letterpaper, 10 pt, conference]{ieeeconf}  

\IEEEoverridecommandlockouts                              

\usepackage{graphics} 
\usepackage{epsfig} 
\usepackage{amsmath} 
\usepackage{amssymb}  
\usepackage{caption}

\usepackage{verbatim}
\usepackage{graphicx} 
\usepackage{epsfig} 
\usepackage{epstopdf}
\usepackage{amsmath} 
\usepackage{amssymb}  
\usepackage{caption}
\usepackage{color}

\makeatletter
\newcommand{\figcaption}{\def\@captype{figure}\caption}
\newcommand{\tabcaption}{\def\@captype{table}\caption}

\newcommand{\tolstrut}{%
	\vrule height\dimexpr\fontcharht\font`\A+.2ex\relax width 0pt\relax
}

\makeatother
\title{\LARGE \bf On Inefficiency of Markowitz-Style Investment\\ Strategies When Drawdown is Important
}

 \author{\large Chung-Han Hsieh$^{1}$ and B. Ross Barmish$^{2}$
 	\thanks{\hskip -10pt ${}^1$Chung-Han Hsieh is a graduate student working towards to his Ph.D. degree in the Department of Electrical and Computer Engineering, University of Wisconsin, Madison, WI 53706. E-mail: hsieh23@wisc.edu.}
 	\thanks{\hskip -10pt ${}^2$B. Ross Barmish is a faculty member in  the Department of Electrical and Computer Engineering, University of Wisconsin, Madison, WI 53706. \mbox{E-mail}: barmish@engr.wisc.edu.}  }

\begin{document}

\maketitle
\thispagestyle{empty}
\pagestyle{empty}

\parindent = 0pt
\begin{abstract}
The focal point of this paper is the issue of \mbox{``drawdown"}  which arises in recursive betting scenarios and related applications in the stock market. Roughly speaking, drawdown is understood to mean drops in wealth over time from peaks to subsequent lows. Motivated by the fact that this issue is of paramount concern to conservative investors, we dispense with the classical variance as the risk metric and work with drawdown and mean return as the risk-reward pair.  In this setting, the main results in this paper address the so-called ``efficiency" of  linear time-invariant (LTI) investment feedback strategies which correspond to Markowitz-style schemes in the finance literature.  Our analysis begins with the following principle which is widely used in finance: Given two investment opportunities, if one
of them has higher risk and lower return, it will be deemed to be inefficient or strictly dominated and generally rejected in the marketplace. In
this framework, with risk-reward pair as described above, our main result is that classical Markowitz-style strategies are inefficient. To establish this, we use a new investment strategy which involves a time-varying linear feedback block~$K(k)$, called the {\it drawdown modulator}. Using this instead of the original LTI feedback block~$K$ in the Markowitz scheme, the desired domination is obtained.  As a bonus, it is also seen that the modulator assures a worst-case level of drawdown protection with probability one.

\end{abstract}


\vspace{3mm}
\section{INTRODUCTION}
\label{INTRODUCTION}
\vspace{-1mm}
The focal point of this paper is the issue of  {\it drawdown}  which arises in recursive betting scenarios and related applications in the stock market; i.e., we consider drops in wealth over time from peaks to subsequent lows. Given that this issue is of paramount concern to conservative investors or bettors,
instead of using the classical variance as the risk metric, we use  the drawdown. Accordingly, our risk-reward pair is obtained using this quantity in combination with the expected return. Beginning with this motivation,
in the sequel, we study  issues of ``efficiency" which arise when linear feedback control strategies are used to adjust the time-varying   investment levels $I(k)$ which are selected at each stage. In the sequel, our understanding is that $I(k)$ denotes either an ``investment" or ``bet." We use these two terms interchangeably.

\vspace{3mm}
The Markowitz and Kelly strategies, in their simplest form, for example see~\cite{Markowitz_1952}-\cite{Kelly_1956}, tell us that the investment $I(k)$ at each stage~$k$ should be ``proportional-to-wealth.'' To be more precise, if~$V(k)$ is the {\it account value} of an investor or bettor at stage~$k$, then such a  strategy is described by time-invariant feedback
$$
I(k) = KV(k)
$$
where the constant $K$  which represents the proportion of the account wagered. We also refer to $I(k)$ above as a \mbox{Markowitz-style} investment function. Typically, when selecting the constant~$K$, we include constraints which we express as~$K \in \mathcal{K}.$ When~$\cal{K}$ includes negative numbers, this is interpreted to mean that short selling is allowed. In this case,~\mbox{$I(k)<0$} indicates that the investor is taking the \mbox{``opposite side"} of the trade or bet being offered. An important example is the case~\mbox{$\mathcal{K} = [-1, 1]$}. In this case,~\mbox{$|I(k)| \leq V(k)$} and that the investment is said to be~\mbox{\it cash-financed}.

\vspace{3mm}
The type of LTI feedback control scheme described above is not only important here but central to our earlier work in~\cite{Barmish_Primbs_2015}-\cite{Malekpour_Barmish_2013}. To see the control-theoretic set-up more clearly, see Figure~\ref{fig:Block_Diagram_KV}. In the figure, the~$X(k)$ are independent and identically distributed random variables representing  {\it return} from the~$k$-th investment~$I(k)$ and the associated gain or loss is $I(k)X(k)$. For the short-selling case, a profit results when~$X(k) < 0$.

\vspace{-2mm}

\begin{center}
	\graphicspath{{Figs/}}
	\includegraphics[scale=0.49]{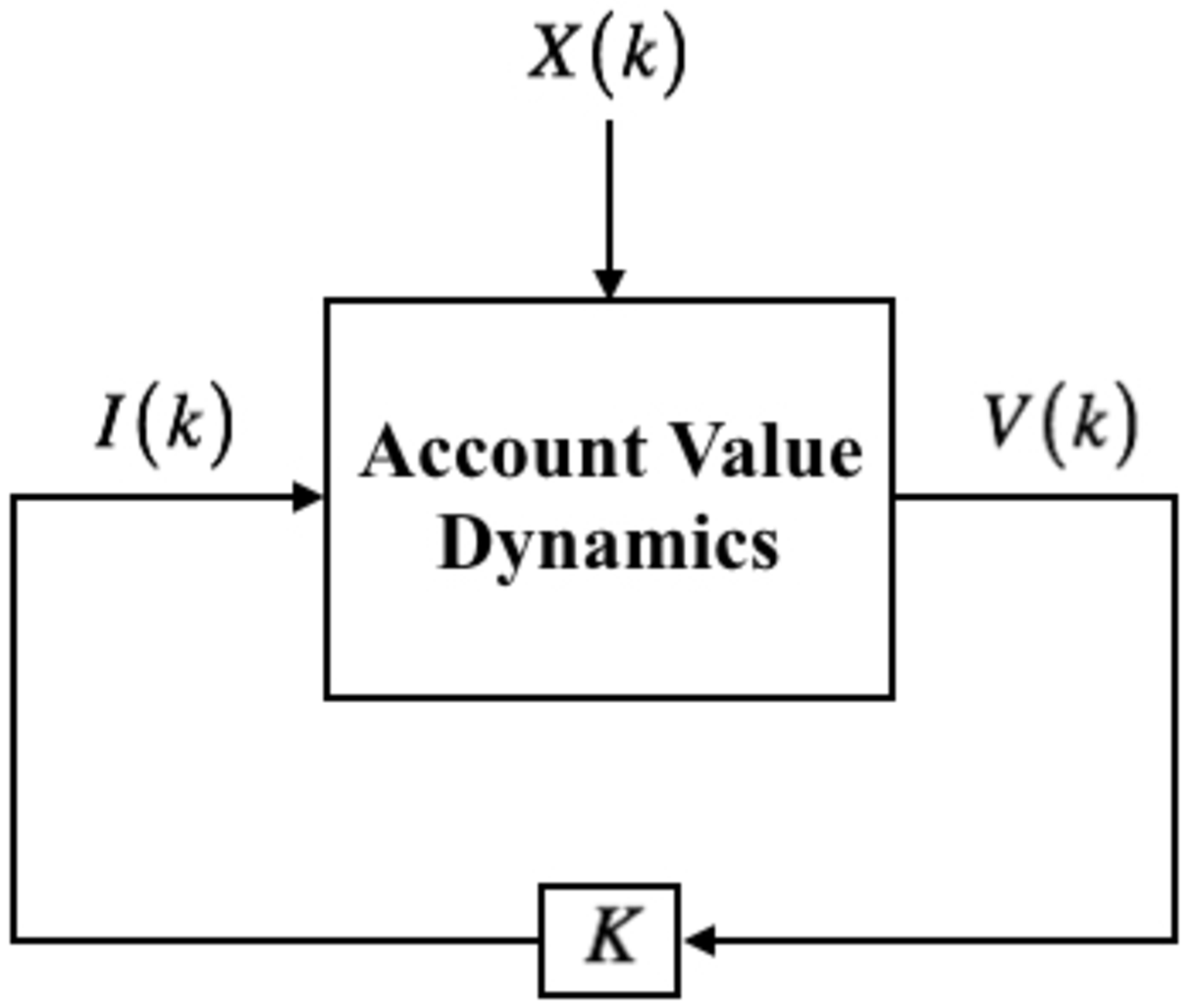}
	\figcaption{Markowitz-Style Gambling Feedback Configuration}
	\label{fig:Block_Diagram_KV}
\end{center}

\vspace{3mm}
{\bf The Notion of Inefficiency}:
The analysis to follow begins with the following principle which is widely used in finance: Given two investment opportunities, if one of them has larger risk and lower return,  it will be  deemed to be {\it inefficient} and generally rejected in the marketplace.   Such an inefficient scheme is  said to be ``strictly dominated." We also refer to a strategy being ``dominated" when the inequalities associated with these conditions are not necessarily strict.
As previously stated, in the literature, the most classical choice for the risk-reward pair is the  variance and expected return;~e.g., see~\cite{Markowitz_1952},~\cite{Markowitz_1959} and \cite{Bodie_Kane_Marcus_2010}. While the use of this pair is quite useful, it relies on an assumption that the returns are normally distributed. Thus, if the distribution of returns is  skewed, then the use of such risk-return metric may be misleading; e.g., see~\mbox{\cite{Malekpour_Barmish_2012}, \cite{Malekpour_Barmish_2013} and \cite{Luenberger_2011}} for more detailed discussion.

\vspace{3mm}
More importantly, as far as this paper is concerned, as previously indicated, instead of using the classical variance as the risk metric when studying efficiency, we use~\mbox{\it drawdown} of wealth which is important from a risk management perspective.  Suffice it to say, the issue of drawdown has received a considerable attention in the finance literature;~e.g., see~\mbox{\cite{Ismail_2004}-\cite{Chekhlov_etal_2005}}. Of these papers, references \cite{Grossman_Zhou_1993} and \cite{Cvitanic_Karatzas_1994} are most relevant. Although their problem setup and assumptions differ from ours, they include one basic idea which  is central to our {\it modulation} controller described below: The investment level is dynamically controlled as a function of~``drawdown to date.'' With the above providing context, our new results on efficiency to follow are based on maximum percentage drawdown and  expected return as the \mbox{risk-reward pair.}

\vspace{3mm}
{\bf Main Results in This Paper}:
To study efficiency, we work with a new feedback-control which generalizes the Markowitz-style investment scheme. This new control includes a constant gain~$\gamma$ and a block $M(\cdot)$ called the~\mbox{\it drawdown modulator} which was  introduced in~\cite{Hsieh_Barmish_2017}; see Figure~\ref{fig:Block_Diagram_ver03}.
With the aid of the modulator block, we show that it is possible to ``dominate" a Markowitz-style strategy; i.e., we obtain the same expected drawdown and higher expected return.  This is made possible by the fact that the modulator $M(\cdot)$ includes memory of~\mbox{$V(0),V(1),...,V(k-1)$} whereas a classical Markowitz-style investment strategy~\mbox{$I(k) = KV(k)$} is memoryless. In addition to our main result on domination described above, as a ``bonus," we also see that the modulator assures a prescribed  level of worst-case drawdown protection which is guaranteed with probability~one.

\vspace{0mm}

\begin{center}
	\graphicspath{{Figs/}}
	\includegraphics[scale=0.24]{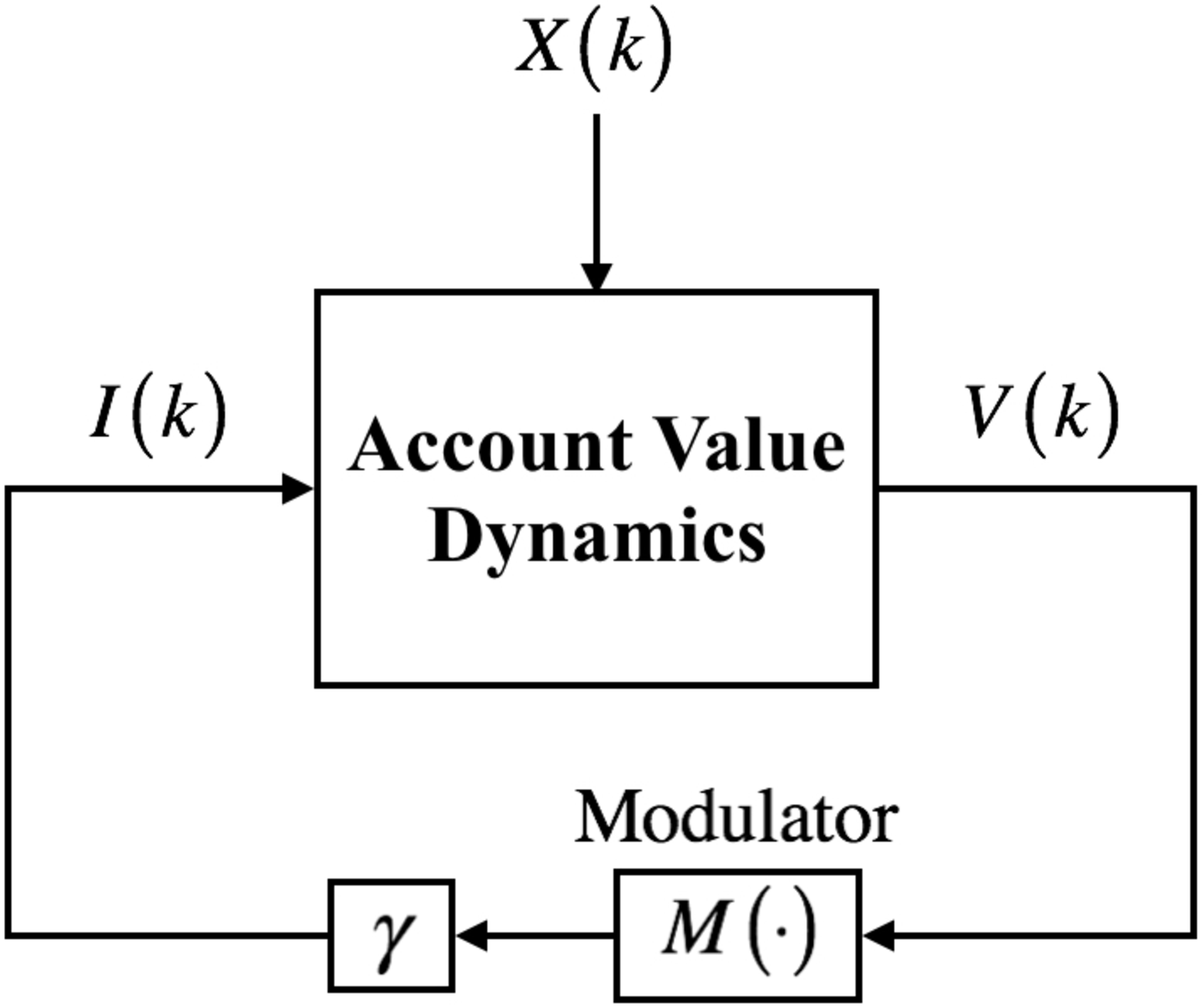}
	\figcaption{Drawdown-Modulated Feedback Configuration}
	\label{fig:Block_Diagram_ver03}
\end{center}

\vspace{2mm}
\section{CLASSICAL DRAWDOWN CONCEPTS}
\label{DRAWDOWN DEFINITION}
\vspace{-1mm}
Consistent with the body of existing literature on drawdown, the definition which we use is as follows:
For~\mbox{$k = 0,1,2,...,N$}, we let $V(k)$ be the corresponding account value. Then, as $k$ evolves, the  {\it percentage drawdown} \mbox{to-date} is defined as
\[
d(k) \doteq \frac{V_{\max}(k) - V(k)}{V_{\max}(k)}
\]
where
\[
V_{\max}(k) \doteq \max_{0 \leq i \leq k} V(i).
\]
This leads to  the  overall {\it maximum percentage drawdown} 
\[
d^* \doteq \max_{0 \leq k \leq N}d(k)
\]
which is central to the analysis to follow.
Note that~\mbox{$ 0 \leq d^* \leq 1.$}
Although not considered in this paper, there is another well-known drawdown-based measure, called the maximum absolute drawdown.
The reader is referred to~\cite{Ismail_2004} and \cite{Hayes} for work on this topic.

\vspace{-2mm}
\begin{center}
	\graphicspath{{Figs/}}
	\includegraphics[scale=0.39]{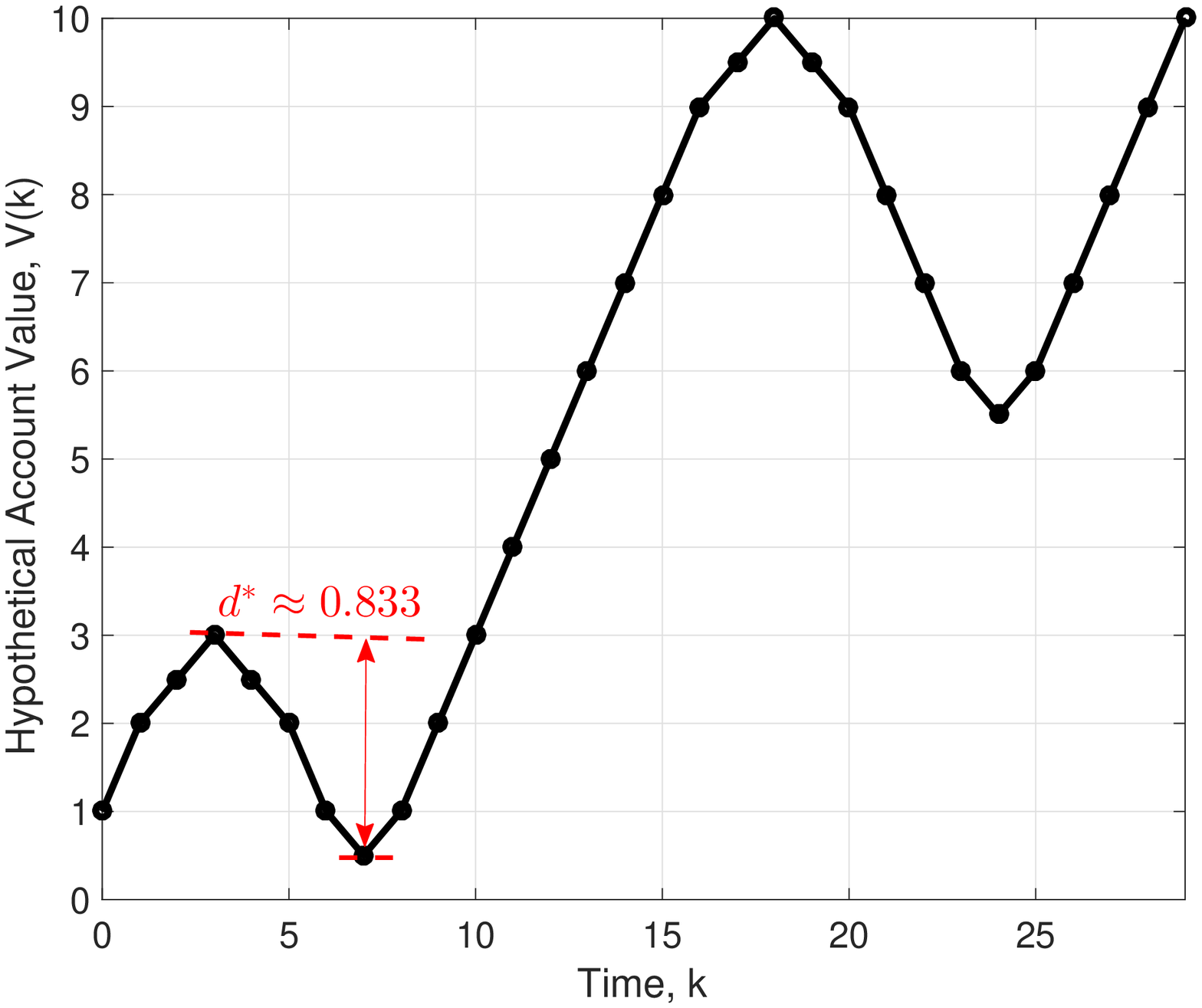}
	\figcaption{Maximum Percentage Drawdown}
	\label{fig:Drawdown_Elaboration}
\end{center}

\vspace{2mm}
{\bf Illustration of Drawdown Definition}:
To further elaborate on the notion of drawdown, we consider an example with a hypothetical account value~$V(k)$ shown in Figure~\ref{fig:Drawdown_Elaboration}.  From the plot,  the overall maximum percentage drawdown
$$
d^* = \frac{3 - 0.5}{3} \approx 0.833
$$
occurs at stage~\mbox{$k=7.$} Note that this maximum percentage drawdown is not necessarily equal to the maximum absolute drawdown which has value $4.5$ and occurs at stage~\mbox{$k=24.$} Percentage drawdown is often used in lieu of absolute drawdown so that the scale of betting is taken into account.

 \vspace{3mm}
\section{INVESTMENT DETAILS AND EFFICIENCY}
\label{KELLY BETTING SCHEMES}
\vspace{-1mm}
In this section, the classical Markowitz-style investment scheme is described in more detail. To begin, for~\mbox{$k=0,1,2,...,N-1$}, we let $X(k)$ be  independent and identically distributed (i.i.d.) random variables which represent {\it returns} for a sequence of sequential bets. We assume that~$
X_{\min} \leq X(k)\leq X_{\max}
$
with $X_{\min}$ and $X_{\max}$ being points in the support, denoted by $\mathcal{X}$, and  satisfying~\mbox{$
X_{\min} < 0<X_{\max}.
$}
Recalling the discussion in Section~\ref{INTRODUCTION}, the $k$-th investment is given by~$
I(k) \doteq KV(k).
$
To assure that the feedback gain~$K$ guarantees $V(k) \geq 0$ for all $k$, we require that 
\[ - \frac{1}{{{X_{\max }}}} \leq K \leq \frac{1}{{\left| {{X_{\min }}} \right|}}\]be satisfied.
 The reader is referred to \cite{Hsieh_Barmish_Gubner_2016} for more details on this condition.
It is also important to note that the~$I(k)<0$ is allowed above; i.e., per Section~\ref{INTRODUCTION}, short selling is admissible.  That is,~\mbox{$I(k)>0$} leads to a profit when $X(k)>0$ and~\mbox{$I(k)<0$}  leads to a profit when~$X(k)<0$. 
Now beginning at initial account value~$V(0)>0$, the evolution to terminal state~$V(N)$ is described sequentially by the recursion
\begin{align*}
V(k+1) 	&= V(k) + I(k) X(k)\\
		&= (1+KX(k))V(k).
\end{align*}
This leads to terminal account value
$$
V\left( N \right) = \prod\limits_{k = 0}^{N - 1} {\left( {1 + K X(k)} \right)V(0)} 
$$
which is useful for calculation of the overall return in the sections to follow.
Although not the focal point of this paper, it is noted that there are many possibilities for selection of the feedback gain $K$ above. Among these possibilities, a popular criterion for selection of $K$ requires maximizing the expected logarithmic growth of wealth; e.g., see~\cite{Kelly_1956}, \cite{Hsieh_Barmisg_2015}, \cite{Hsieh_Barmish_Gubner_2016},~\cite{MacLean_Thorp_Ziemba_2011} and~\cite{Thorp_2006}.

 \vspace{3mm}
 {\bf Efficiency Considerations}:
 The question now arises  regarding the extent to which a Markowitz-style investment scheme is efficient.  Indeed, against any sample path $X(k)$, we let~$R_K$ denote the {\it overall return}; i.e.,
\begin{align*}
 R_K &\doteq \frac{V(N) - V(0)}{V(0)}
  = \prod\limits_{k = 0}^{N - 1} {\left( {1 + KX\left( k \right)} \right)}  - 1
 \end{align*}
  and along the path, we obtain 
  $$
  d_K^* \doteq \max_{0 \leq k \leq N} d_K(k)
  $$ 
 which is the maximum percentage drawdown. Note that the subscript $K$ in $R_K$ and $d_K^*$ is used to emphasize the dependence on the feedback gain $K$. Now, to study efficiency,  in the sections to follow, we use the expected values of $R_K$ and $d_K^*$ below. Using the shorthand 
 $$
 \overline{\tolstrut R}_K \doteq \mathbb{E}[R_K]
 $$ 
 and
 $$
 \overline{\tolstrut d}^{_*}_K \doteq \mathbb{E}[d_K^*]
 $$ 
 to denote these quantities,  we obtain the {\it attainable risk-return performance curve} in the plane as
 $$
 \{ {( \overline{\tolstrut R}_K,\overline{\tolstrut d}^{_*}_K ):K \in \mathcal{K} } \}.
$$
In addition, recalling that~$X(k)$ are independent and identically distributed, letting~\mbox{$\mu \doteq \mathbb{E}[X(k)]$} and using the formula for $R_K$ above, we obtain
 \begin{eqnarray*}
 \overline{\tolstrut R}_K  &=& \mathbb{E}\left[ {\prod\limits_{k = 0}^{N-1} {\left( {1 + KX\left( k \right)} \right)} } \right] - 1 \hfill \\[2pt]
 &=& {\left( {1 + K\mathbb{E}\left[ {X\left( k \right)} \right]} \right)^N} - 1 \\[2pt]
 &=& {\left( {1 + K\mu} \right)^N} - 1 .
 \end{eqnarray*}
As far as calculation of~$\overline{d}_K^{_*}$ is concerned, except for small values of $N$, Monte-Carlo simulations are used to calculate this quantity; see Section~\ref{EXAMPLE DEMONSTRATING INEFFICIENCY} for more detail. Further to~$\overline{d}_K^{_*}$, it is straightforward to see that  the worst-case percentage drawdown is
\[
 d^* \doteq 1- (1 - |K| \max \{|X_{\min}|,X_{\max}\})^N
\]
which is much less useful than $\overline{\tolstrut d}_K^{_*}$ since it corresponds to losing every bet and typically has very low probability.
For example, for a simple even-money payoff coin-flipping gamble with~\mbox{$N=10$} and probability of heads~\mbox{$P(X(k) = 1) = p =  0.6$}, the celebrated Kelly optimum~$K= 2p-1 = 0.2$ obtained in papers such as~\cite{Kelly_1956} and~\cite{Thorp_2006}, leads to
$
d^* = 1-(1 - 0.2)^{10} \approx 0.89
$
which corresponds to~$89\%$ as the worst-case.

\vspace{3mm}
\section{DRAWDOWN MODULATION}
\label{DRAWDOWN MODULATION}
\vspace{-1mm}
The starting point for this section is the fact that the classical Markowitz-style investment strategy above is \mbox{``memoryless."} That is, at stage $k$, the investment $I(k)$ does not depend on~$V(0),V(1),...,V(k-1)$. We now describe how the inclusion of a ``modulator block," first introduced in~\cite{Hsieh_Barmish_2017},  can be used to improve performance when the risk metric is percentage drawdown. As shown below, this is a time-varying generalization of the linear feedback scheme~$I(k) = KV(k)$.  To begin, given a prescribed constant~\mbox{  $0 <d_{\max}<1$} which represents the~\mbox{\it maximum allowable percentage drawdown}, the following lemma plays an important in our theory. It provides a necessary and sufficient condition for {\it any} investment strategy~$I(k)$ assuring that~\mbox{$
d(k) \leq d_{\max}
$}
with probability one. For the sake of self-containment of this paper, we include the proof from~\cite{Hsieh_Barmish_2017} here.

\vspace{3mm}
{\bf The Drawdown Modulation Lemma}: {\it An investment function $I(\cdot)$ guarantees maximum acceptable drawdown level~$d_{\max}$ or less with probability one if and only if the condition, 
	\[ 
- \frac{{{d_{\max }} - d(k)}}{{\left( {1 - d\left( k \right)} \right){X_{\max }}}}V(k) \leq I(k) \leq \frac{{{d_{\max }} - d(k)}}{{\left( {1 - d\left( k \right)} \right)\left| {{X_{\min }}} \right|}}V(k)
\]
	is satisfied along all sample pathes.}

\vspace{3mm}
{\bf Proof}: To prove necessity, assuming that
$d(k) \le d_{\max}$  for all~$k$ with probability one, we must show the required condition on~$I(k)$ holds along all sample pathes. Indeed, letting~$k$ be given, since both \mbox{$d(k) \le d_{\max}$} and $d(k+1) \le d_{\max}$ with probability one, we claim this forces the required inequalities on $I(k)$. Without loss of generality, we provide a proof of the rightmost inequality for the case~{$I(k) \ge 0$} and note that a nearly identical proof is used for~{$I(k) < 0$}.  Indeed,  using the fact that~$X_{\min}$ is in the support~$\cal X$,  
there exists a suitably small neighborhood of~$X_{\min}$, call it $\mathcal{N}(X_{\min})$, such that
\[
P \left (X(k) \in \mathcal{N}(X_{\min}) \right) >0.
\]
Thus, given any arbitrarily small~$\varepsilon >0$, there exists some point $X_{\varepsilon}(k) < 0$ such that $
X_{\varepsilon}(k) \in \mathcal{N}_\varepsilon(X_{\min}) ,
$
$|X_{\min} - X_\varepsilon(k)| < \varepsilon $ and leading to
associated realizable loss $I(k)|X_
\varepsilon (k)|$.
Noting that~$V_{\max}(k+1) = V_{\max}(k)$ and
\begin{align*}
d(k + 1)= d(k) + \frac{{\left| {{X_\varepsilon (k) }} \right|I(k)}}{{{V_{\max }}(k)}} \leq d_{\max}
\end{align*}
we now substitute
\[{V_{\max }}(k) = \frac{{V\left( k \right)}}{{1 - d\left( k \right)}} >0
\]
into the inequality above and noting that $|X_\varepsilon(k)| \to |X_{\min}|$ as~$\varepsilon \to 0$, we obtain
\[
I(k) \le \frac{{{d_{\max }} - d(k)}}{{\left( {1 - d\left( k \right)} \right)\left| {{X_{\min }}} \right|}}V(k).\]
To prove sufficiency,  we assume that the condition on  $I(k)$ holds along all sample pathes. We must show~\mbox{$d(k) \le d_{\max}$}  for all~$k$ with probability one.  Proceeding by induction,  for~$k=0$, we trivially have~\mbox{$d(0)=0 \leq d_{\max}$} with probability one.  To complete the inductive argument, we assume that~$d(k) \le d_{\max}$ with probability one, and  must  show~\mbox{$d(k+1) \le d_{\max}$} with probability one. Without loss of generality, we again provide a proof for the case~{$I(k) \ge 0$} and note that a nearly identical proof is used for~{$I(k) < 0$}. Now, by noting that
\begin{align*}
d\left( {k + 1} \right) 
&= 1 - \frac{{V\left( {k + 1} \right)}}{{{V_{\max }}\left( {k + 1} \right)}},
\end{align*}
and $V_{\max}(k) \leq V_{\max}(k+1)$ with probability one, we split the argument into two cases: If $V_{\max}(k) < V_{\max}(k+1)$, then $V_{\max}(k+1) = V(k+1).$ Thus, we have $d(k+1)=0 \leq d_{\max}.$ On the other hand, if $V_{\max}(k) = V_{\max}(k+1)$, with the aid of the dynamics of account value, we have
\begin{align*}
d\left( {k + 1} \right) 
&= 1 - \frac{{V\left( k \right) + I\left( k \right)X\left( k \right)}}{{{V_{\max }}\left( k \right)}}\\
& \le 1 - \frac{{V\left( k \right) - I\left( k \right)\left| {{X_{\min }}} \right|}}{{{V_{\max }}\left( k \right)}}
\end{align*}
Using the rightmost inequality condition on $I(k)$,  we obtain $ d\left( {k + 1} \right) \le {d_{\max }}$
which completes the proof. $\;\;\;\;\; \square$

\vspace{3mm}
{\bf Drawdown-Modulated Feedback Control}:
Motivated by the lemma above, we now  consider a time-varying feedback control of the form
 $$
I(k) =  K(k)V(k)
$$
with $
K(k)  = \gamma M(k)
$
where
\[
M(k) \doteq \frac{d_{\max} - d(k)}{1 - d(k)}
\]
and
\[ 
- \frac{1}{{{X_{\max }}}} \leq \gamma  \leq \frac{1}{{|{X_{\min }}|}}.
\]  

Note that $0 \leq M(k) \leq d_{\max}$. In the sequel, the constraint above on~$\gamma$ is denoted by writing $\gamma \in \Gamma$.  
In the next section, we see how the two design variables~\mbox{$d_{\max} \in (0,1)$} and~$\gamma \in \Gamma $  are selected by the designer when we study the efficiency~issue.



\vspace{3mm}
\section{THE DOMINATION LEMMA}
\label{THE DOMINATION LEMMA}
\vspace{-1mm}
We now show  that with drawdown-modulated feedback, it  is possible to ``dominate" a Markowitz-style strategy;~i.e., it leads to the same expected drawdown and possibly higher expected return. As a bonus, as previously stated, we also see that the modulator assures a pre-specified worst-case level of drawdown protection with probability one.
 
\vspace{3mm}
{\bf Attainable Risk-Return Performance}:
Henceforth, we use notation
\[
\mathcal{M} \doteq (\gamma, d_{\max}) \in \Gamma \times (0,1)
\]
to denote an admissible drawdown modulation pair. Then, the associated return and maximum percentage drawdown is denoted by $R_\mathcal{M}$ and $d_\mathcal{M}^*$, respectively. Hence, for the expected return and expected maximum drawdown, we write
$$
\overline{\tolstrut R}_{\mathcal{M}} \doteq \mathbb{E}[R_\mathcal{M}]
$$ 
and 
$$
\overline{\tolstrut d}_{\mathcal{M}}^{_*} \doteq \mathbb{E}[d_\mathcal{M}^*].
$$
This leads to the {\it attainable risk-return performance set} in the plane described by
\[
\left\{ {\left( {{{\overline {\tolstrut R}}_\mathcal{M}},\overline{\tolstrut d}_\mathcal{M}^{_*}} \right):\mathcal{M} \doteq (\gamma, d_{\max}) \in \Gamma \times (0,1) } \right\}.
\]
To obtain points in the set above, we use an idea which is found in the celebrated Markowitz risk-return theory in finance;~e.g., see~\cite{Markowitz_1952} and \cite{Markowitz_1959}. That is, given any target level of expected drawdown, call it $\widehat{d}$, we seek an admissible pair~\mbox{$\mathcal{M} \doteq (\gamma, d_{\max}) \in \Gamma \times (0,1)$} maximizing
$
\overline{\tolstrut R}_\mathcal{M}
$
subject to the constraint~$\overline{\tolstrut d}_\mathcal{M}^{_*} = \widehat{d}.$ In our case, this is found by solving a two-dimensional optimization over the rectangle constraining~$\gamma$ and~$d_{\max}$~above. We are now prepared to address the issue of domination.

\vspace{3mm}
{\bf The Domination Lemma}:
{\it For any admissible~$K \in \mathcal{K}$, there exists a drawdown modulator pair~\mbox{$
\mathcal{M} = (\gamma, d_{\max} ) $} such that $$
\overline{\tolstrut R}_\mathcal{M} \geq \overline{\tolstrut R}_K
$$
and
$$
\overline{\tolstrut d}_\mathcal{M}^{_*} = \overline{\tolstrut d}_K^{_*}.
$$
}

\vspace{-2mm}
{\bf Proof}: To begin, taking the target level of drawdown~$\widehat{d} \doteq \overline{d}_K^*$, we must show that there is an admissible pair~\mbox{$\mathcal{M} = (\gamma,d_{\max})$} which leads to
$\overline{\tolstrut d}_\mathcal{M}^{_*} = \widehat{d}$
and~\mbox{$\overline{\tolstrut R}_\mathcal{M} \geq \overline{\tolstrut R}_K.$}
Indeed, taking~\mbox{$\gamma = K$} and letting~$d_{\max} \to 1$, we first replicate the performance of  Markowitz-style investment scheme;~i.e., we obtain~\mbox{$
\overline{\tolstrut d}_\mathcal{M}^{_*} = \overline{\tolstrut d}_K^{_*}
$}
and~\mbox{$
\overline{\tolstrut R}_\mathcal{M} = \overline{\tolstrut R}_K.
$} 
Hence the maximization of $\overline{\tolstrut R}_\mathcal{M}$ over all admissible $\mathcal{M} \in   \Gamma \times (0,1)$ with constraint~$\overline{\tolstrut d}_\mathcal{M}^{_*} = \overline{\tolstrut d}_K^{_*}$ must be at least as large as $\overline{\tolstrut R}_K.$\;~$\square$

\vspace{3mm}
{\bf Remarks}: Note that the Markowitz-style strategy can be viewed as a subclass of drawdown-modulated class  obtained with~\mbox{$\gamma = K$} and~$d_{\max} \to 1$. 
Furthermore, as demonstrated in the Section~\ref{EXAMPLE DEMONSTRATING INEFFICIENCY}, it is typically the case that the inequality in the lemma above is ``strict." In other words, the Markowitz-style investment scheme may be ``strictly dominated" by a strategy in the modulator class.

\vspace{3mm}
 \section{ILLUSTRATIVE EXAMPLES}
\label{EXAMPLE DEMONSTRATING INEFFICIENCY}
\vspace{-1mm}
In many applications, the broker's constraint on leverage forces the satisfaction of the cash-financing condition~$|I(k)| \leq V(k)$; i.e., for drawdown modulated feedback, to guarantee this condition is satisfied, the constraint on~$\gamma$ described in Section~\ref{DRAWDOWN MODULATION} is augmented to include~\mbox{$
|\gamma| \; M(k) \leq 1 .
$}
Similarly, for a Markowitz-style investment strategy, to guarantee the cash financing condition, we augment the constraint on $K$ to include $|K| \leq 1.$ In the examples to follow the constraint which we impose on the Markowitz-style investment is also used for the modulation scheme.

\vspace{3mm}
We now illustrate use of our result on domination via examples.
We begin with the simple case when $N = 2$ where calculations can be carried out in closed form. Then we study the more general scenario with $N>2$ where Monte Carlo simulation is used. Indeed, for $N=2$, we consider a single coin-flipping gamble having even-money payoff described by independent and identically distributed random variables~\mbox{$X(k) \in \{-1, 1\}$} and~\mbox{$P(X(k)= 1) \doteq p >1/2$}.  
 
\vspace{3mm}
{\bf Reward-Risk Calculations for Both Schemes}:
Now, beginning with~\mbox{$ \mu = \mathbb{E}[X(k)] = 2p-1$}, for the Markowitz-style betting strategy with parameter $ K > 0$, the associated expected return is readily calculated to be
 \begin{align*}
 \overline{\tolstrut R}_{K} 
 &= {\left( {1 + K\left( {2p - 1} \right)} \right)^2} - 1.
 \end{align*} 
 and the expected maximum percentage drawdown, found by a straightforward calculation is given by
 \begin{align*}
\overline{\tolstrut d}^{_*}_{K} &=K(1 - p)\left( {2 - K + Kp} \right).
 \end{align*} 
For drawdown modulator pair $\mathcal{M} \doteq (\gamma, d_{\max})$, a lengthy but straightforward computation leads to  expected return and expected maximum percentage drawdown given by
 \begin{align*}
 \overline{\tolstrut R}_\mathcal{M} 
 &= \gamma {d_{\max }}(2p - 1)(\gamma {d_{\max }}p + \gamma p - \gamma  + 2)
 \end{align*} 
 and
 \begin{align*}
 \overline{\tolstrut d}_\mathcal{M}^{_*}  & = \gamma {d_{\max }}\left( {1 - p} \right)\left( {2 - \gamma  + \gamma p} \right). 
 \end{align*}

\vspace{3mm}
{\bf Demonstration of Strict Domination}: 
Now, we establish  ``strict domination" using drawdown-modulated feedback strategy. That is, for any $0<K < 1$,  we prove that there exists a modulator $\mathcal{M} = (\gamma, d_{\max})$ such that 
	$\overline{\tolstrut R}_K < \overline{\tolstrut R}_\mathcal{M}^{_*}
	$
	and
	$
	\overline{\tolstrut d}_\mathcal{M}^{_*} = \overline{\tolstrut d}_K^{_*}.
	$
Indeed, to prove this, it is sufficient to take~$\gamma =1$~and
\[
{d_{\max }} = \frac{{K\left( {2 - K + Kp} \right)}}{{1 + p}}
\]
which is obtained by setting $\overline{\tolstrut d}_\mathcal{M}^{_*} = \overline{\tolstrut d}_K^{_*}$ above. It is readily verified that~\mbox{$ 0 < d_{\max }<1$} and by substitution of $d_{\max}$ and~$\gamma$ into~$\overline{\tolstrut R}_\mathcal{M}$,  after a lengthy but straightforward calculation, we obtain
\[
{ \overline{\tolstrut R}_\mathcal{M}} = \frac{{K(2p - 1)(2 - K + Kp)f\left( {K,p} \right)}}{{{{(1 + p)}^2}}}
\]
where $
f(K,p) \doteq 2Kp - {K^2}p + {K^2}{p^2} + {p^2} + 2p + 1.
$
Now, to establish the desired domination, we now claim that~\mbox{$
\overline{\tolstrut R}_\mathcal{M} > \overline{R}_K.
$ }
To prove this, we show that the difference between left and right hand sides above is positive. Indeed, via a  lengthy but straightforward calculation, we obtain
{\small
\begin{align*}
\overline R_\mathcal{M} - \overline R_K^{} 
& = \frac{{{K^2}(1 - K)(1 - p)p(2p - 1)(3 + p + Kp - K)}}{{{{(1 + p)}^2}}}.
\end{align*}
}Noting that $ 0 < K < 1$ and $p > 1/2$ above, it is immediate that  both numerator and denominator for the difference described above are strictly positive.
Thus,~\mbox{$\overline{\tolstrut R}_{\mathcal M} > \overline{\tolstrut R}_K.$} To complete this analysis, in Figure~\ref{fig:Domination_Plot_for_Coin_N_2.eps} we provide a plot which shows the degree of  the strict domination in the difference  based on our calculation of  $\overline{\tolstrut R}_{\mathcal M} - \overline{R}_K$ above.

\vspace{-2mm}
\begin{center}
	\graphicspath{{Figs/}}
	\includegraphics[scale=0.45]{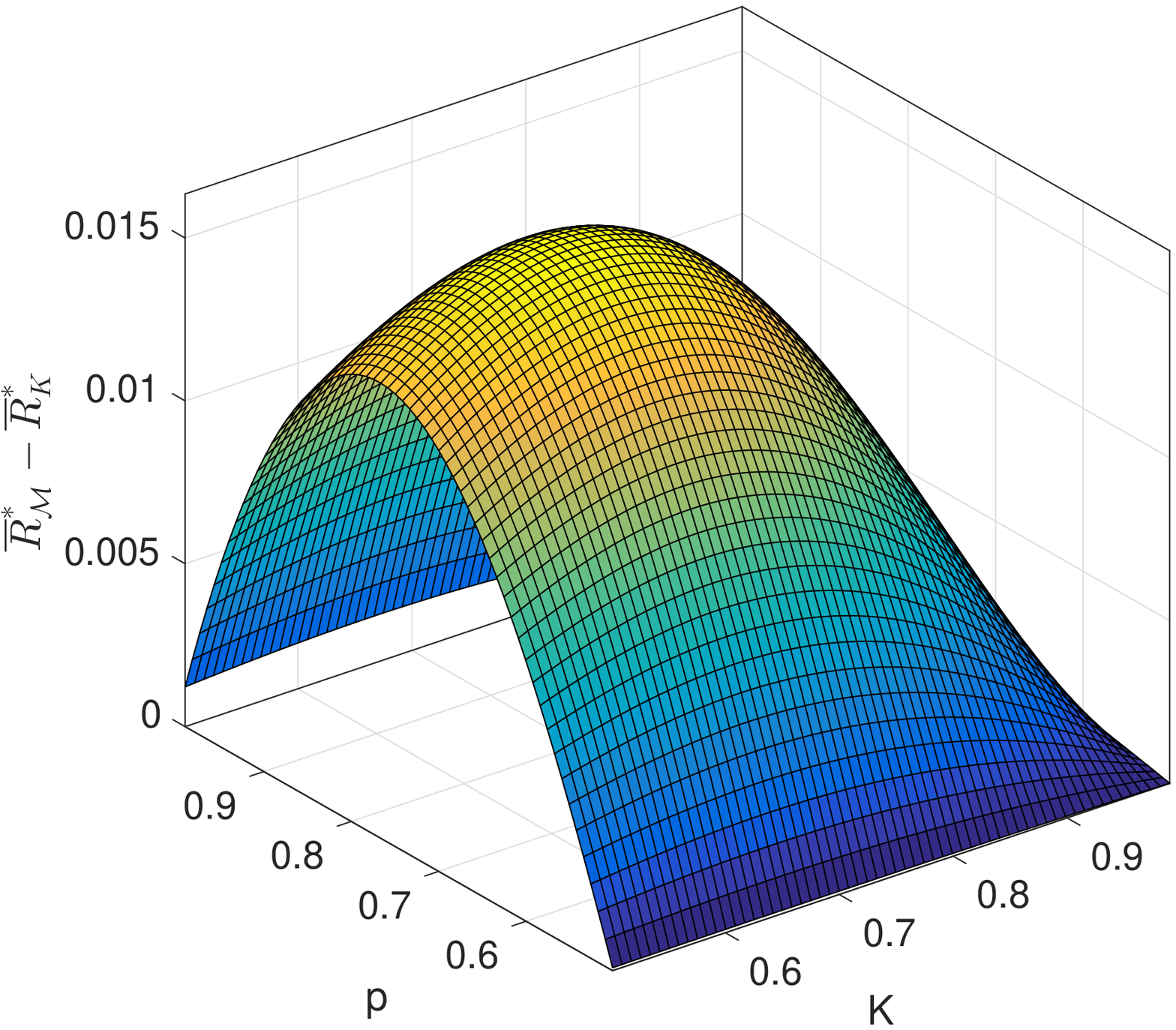}
	\figcaption{Degree of Strict Domination in Expected Return}
	\label{fig:Domination_Plot_for_Coin_N_2.eps}
\end{center}  

 
\vspace{2mm}
{\bf Example of Inefficiency with Larger $N$}:
Again, we consider a single coin-flipping scenario with even-money payoff described by independent and identically distributed random variables \mbox{$X(k) \in \{-1/30, 1/30\}$} and \mbox{$P(X(k)= 1/30)= 0.6$}
with corresponding mean \mbox{$
\mu = \mathbb{E}[X(k)] =  1/{150}.
$}
We choose~$N=252$ and view this as  a trading problem for a binomial stock-price model over one year with daily returns varies around~$ \pm 3.3\%$ corresponding to $X(k) = 1/30$ above.  Note that this scenario is more biased on~$X(k)=1/30$.  Hence, we  study efficiency for the case when~$0 \leq K \leq 1$. 

\vspace{3mm}
{\bf Demonstration of Inefficiency}: For the Markowitz scheme, we first obtain the expected return
\begin{align*}
  \overline{\tolstrut R}_K  
   &= {\left( {1 + \frac{K}{{150}}} \right)^{252}} - 1.
\end{align*}
As far as the expected maximum percentage drawdown~$\overline{d}_K^{_*} $ is concerned, this quantity is computed via performing a large number of Monte-Carlo simulations.
Our finding is that for~$0 \leq K \leq 1,$ we have $
\overline{\tolstrut d}_K^{_*} \approx 0.25 K.
$
%
%
%
For the drawdown-modulated feedback with the cash-financing condition imposed, to certify inefficiency, we proceed as follows: As previously discussed in Section~\ref{DRAWDOWN MODULATION}, for a given target level of drawdown~$\widehat{d} \in (0,1)$, we seek to find a pair~\mbox{$\mathcal{M} = (\gamma^*, d_{\max})$} maximizing~$
\overline{\tolstrut R}_\mathcal{M}
$
subject to~$
\overline{\tolstrut d}_\mathcal{M}^{_*} = \widehat{d}.
$
This two-parameter optimization is solved with a large Monte-Carlo simulation. Then, letting~$\overline{\tolstrut R}_\mathcal{M}^{_*}(\widehat{d})$  denote the approximate optimal value obtained, we generate the dotted line in the Figure~\ref{fig:Efficiency_Plot.eps}. 
%
We see that for any given risk level, the drawdown-modulated feedback leads to a certifiably higher expected return than the Markowitz-style investment scheme. 
In other words, the Markowitz-style investment scheme is ``strictly dominated" as seen in the figure.

\vspace{-5mm}

\begin{center}
	\graphicspath{{Figs/}}
	\includegraphics[scale=0.45]{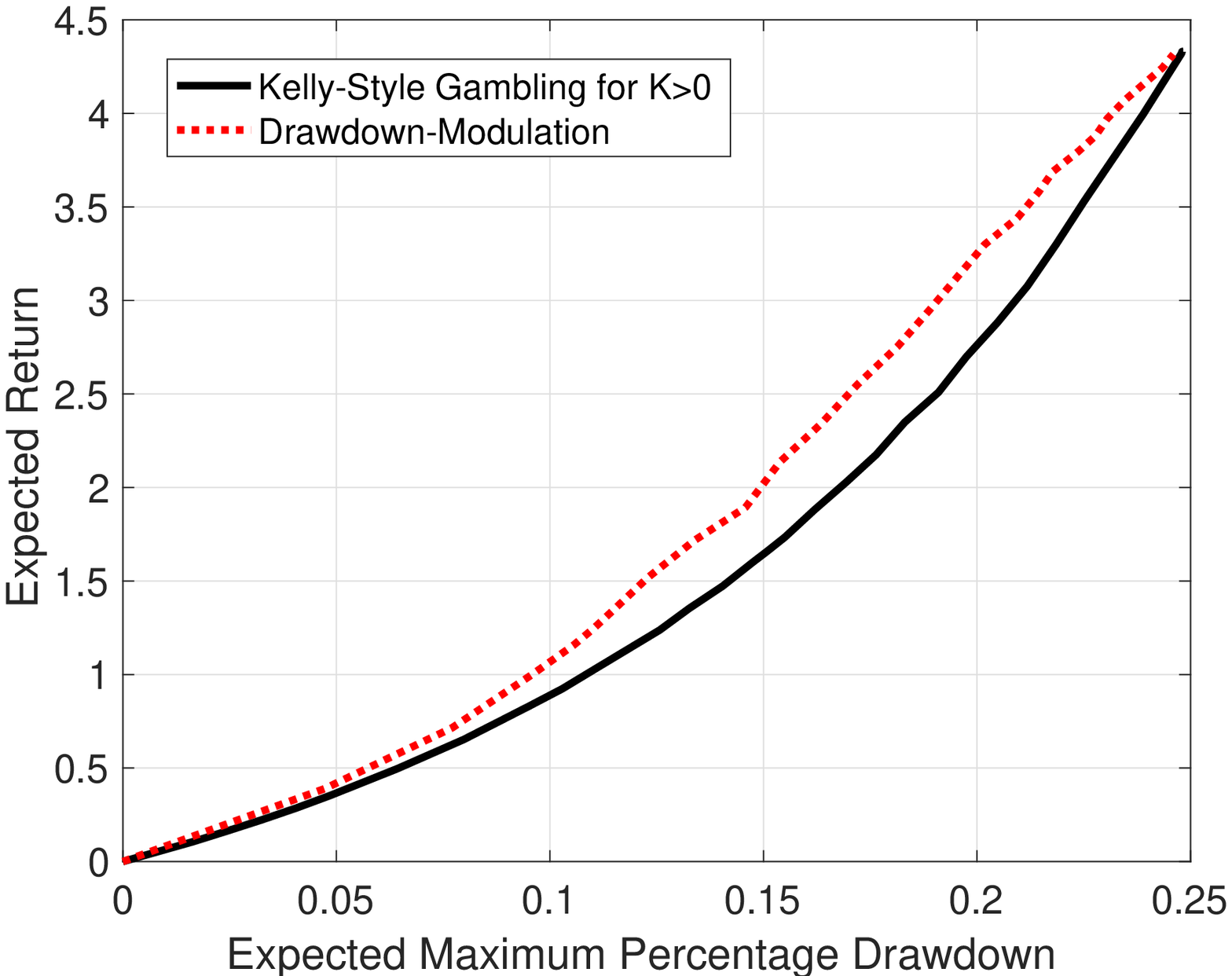}
	\figcaption{Risk-Reward Efficiency Plot for $N=252$}
	\label{fig:Efficiency_Plot.eps}
\end{center}

\vspace{1mm}
\section{CONCLUSION AND FUTURE WORK}
\vspace{-1mm}
In this paper, using  expected maximum percentage drawdown and mean return as the risk-reward pair, we demonstrated inefficiency of Markowitz-style investment schemes. This was accomplished using our so-called drawdown-modulated feedback control. 
In addition, as a bonus, this controller was seen to guarantee a prescribed level of drawdown protection with probability one. By way of extending the results given in this paper, it is interesting to note that a drawdown-modulated controller can be used to obtain  very similar domination results for other return metrics as well. For example, if $\overline{\tolstrut R}_K$ is replaced by the expected logarithmic growth~$\mathbb{E}[\log(V(N)/V(0)])$, which is central to papers such as~\cite{Kelly_1956}, \cite{Hsieh_Barmisg_2015}, \cite{Hsieh_Barmish_Gubner_2016},  \cite{MacLean_Thorp_Ziemba_2011} and \cite{Thorp_2006},  performance comparisons are obtained which are very similar to that given in Figure~\ref{fig:Efficiency_Plot.eps} result.


\vspace{3mm}
Regarding further research on efficiency issues, one obvious extension would be to consider a portfolio case which involves many correlated random variables;~i.e., we take~$X(k)$ to be a vector rather than the scalar considered here. When~$X(k)$ has dimension $n$ which is large, finding the attainable performance curve, often called the~\mbox{\it efficient frontier}, may require an  algorithm aimed at dealing with high computational complexity.
Another interesting problem for future research is motivated by the fact that the feedback gain for our drawdown-modulated feedback scheme we used is simply a pure gain~$\gamma$. It may prove to be the case that a time-varying feedback gain $\gamma(k)$ may lead to superior performance in the risk-reward efficiency framework. Finally, as seen in Section~\ref{THE DOMINATION LEMMA},  the  lemma does not guarantee ``strict" domination. An interesting extension of this work  would be to provide  conditions under which strictness can be guaranteed.




\end{document}